\documentclass[prl, twocolumn, superscriptaddress]{revtex4-1}
\usepackage{amsmath}
\usepackage[english]{babel}
\usepackage[pdftex]{graphicx}
\usepackage{color}
\usepackage{ulem}
\usepackage{comment}

\usepackage[sort&compress]{natbib}
\usepackage{hyperref}

\graphicspath{{images/}{./}}

\usepackage{amsmath}
\usepackage{amssymb}
\usepackage{mathtools}
\usepackage{braket}

\renewcommand{\vec}[1]{\ensuremath{\boldsymbol{#1}}}

\begin{document}
\title{Dual energy-differentiated topological transition in artificial red phosphorus chains}

\author{Vít Jakubský}
\email{jakubsky@ujf.cas.cz}
\affiliation{Nuclear Physics Institute, Czech Academy of Science, 250 68 \v Re\v z, Czech Republic}

\author{B. Manjarrez-Monta\~nez}
\email{manjarrez.br@icf.unam.mx}
\affiliation{Instituto de Ciencias F\'isicas, Universidad Nacional Aut\'onoma de M\'exico, Cuernavaca, M\'exico}

\author{Rafael A. M\'endez-S\'anchez}
\email{mendez@icf.unam.mx}
\affiliation{Instituto de Ciencias F\'isicas, Universidad Nacional Aut\'onoma de M\'exico, Cuernavaca, M\'exico}

\author{Yonatan Betancur-Ocampo}
\email{ybetancur@fisica.unam.mx}
\affiliation{Instituto de F\'isica, Universidad Nacional Aut\'onoma de M\'exico, Ciudad de México, Mexico}

\begin{abstract}
We investigate the spectral and transport properties of an atomic chain of red phosphorus. We reveal the separation of flat-band states from the rest of the system and calculate its energy bands analytically. The topological properties of the system are established through the evaluation of the Berry (Zak) phase of the energy bands, revealing nontrivial topology. The Berry phase depends on the relative strength of the hopping parameters and exhibits dual energy-dependent topological phase transitions. Remarkably, the emergence of inert band edges provides a direct spectral signature of these transitions, acting as energy-resolved indicators of the redistribution of topological charge between bands. The existence of the associated edge states is proved numerically for finite lattices. The theoretical predictions, particularly the band structure and the existence of edge states, are further confirmed by numerical simulations of red phosphorus through a phononic lattice in the form of a highly structured aluminum plate. 
\end{abstract} 
\maketitle

\section{Introduction}
The study of topological features in one-dimensional chains has been of enormous interest in research due to the potential for technological applications, including topological qubits in quantum computing \cite{Kitaev2003,Lutchyn2010,Oreg2010,Tsintzis2024}. From the theoretical point of view, one-dimensional chains are the simplest systems to understand difficult concepts in topological insulators. Moreover, artificial crystals as acoustic~\cite{Sigalas1998,Escalante2013,Li2018a,Martinez-EsquivelEtAl,Wu2021,Coutant2021,Wang2022,Liao2022,DalPoggetto2024,Ongaro2025,Tang2025}, photonic \cite{AbarcaRamirez2025,Bayindir2000,Bittner2010,Verbin2013,Mukherjee2015,Dietz2018,Li2019,Setare2019,Xue2021,CaceresAravena2022,Francis2022,Liu2022,Tang2022}, topolectric circuits \cite{Guo2025a}, or phononic metamaterials \cite{BetancurOcampo2024,ManjarrezMontanez2025} are ideal platforms for testing phenomena predicted from tight-binding models. The main advantage arises from the versatility to manipulate straightforward evanescent couplings and geometry in resonator arrays.

There are many important examples of one-dimensional chains with experimental realizations. The most famous is the Su-Schrieffer-Heeger model, which embodies the topological phase transition in trans-polyacetylene \cite{Fradkin1983,Li2014,Su1979,Heeger1981,Heeger1988,Obana2019}. 
This model predicted a topological phase transition when one of the hopping parameters is increased \cite{Asboth2016}. Such a topological phase has been confirmed in multiple experimental realizations, such as semiconductor quantum dots~\cite{Kiczynski2022}, topolectrical circuits~\cite{Lee2018,Wang2019,Guo2025a}, superconducting circuits \cite{Splitthoff2024}, cold atoms~\cite{Meier2016}, Rydberg atoms~\cite{Kanungo2022}, acoustic chains \cite{Li2018a,Coutant2021}, polariton lattices~\cite{Pernet2022}, water waves, elastic strings and  metamaterials~\cite{Anglart2025,Thatcher2022,Wang2024}, photonic lattices and microwave resonators~\cite{CaceresAravena2022,Poli2015}. 
Based on this model, similar and extended SSH models have been proposed, for instance, non-hermitian SSH chains~\cite{Lieu2018,Fan2022,Liu2022a,ManyManda2024}, two-dimensional SSH lattices \cite{Kim2020,Xu2020,Francis2022,Li2022,Chen2022a}, bearded SSH chains \cite{CaceresAravena2022,BetancurOcampo2024a,Jakubsky2024,Jakubsky2025}, and a twofold topological phase transition in cis-polyacetylene \cite{BetancurOcampo2024,AbarcaRamirez2025}. Each unusual version of the SSH model offers exotic phenomena and represents a significant advance in the development and understanding of topological insulator theory. 

In this paper, an artificial crystalline version inspired by the red phosphorus molecule is proposed. We present a tight-binding model for both finite and infinite periodic chains to describe their electronic properties and topological features, where two non-trivial phases occurring at different location of the energy spectrum are found. Calculation of Berry phase and edge states for the finite chain confirms the emergence of these topological phases. An artificial chain based on a coupled-resonator phononic metamaterial is proposed, where the elastic wave equation is solved numerically with the finite element method. In this system, the aluminium octagon resonator plays the role of a meta-atom while waveguides couple evanescently to the resonators. The comparison of tight-binding energy bands and phononic diagram shows an excellent agreement. Moreover, the prediction of edge states from the model is confirmed by the normal vibration modes of the artificial chain. 

The article is organized as follows. In the first section, we introduce the red-phosphorus chain within the tight-binding framework. Then, the decoupling of the flat-band dynamics and equivalence of the system to an SSH-like trimer chain are addressed in the second section. Analytical expressions of the Bloch Hamiltonian, band structure, and Berry phase are calculated to topologically characterize the chain in the subsequent sections. We focus on the existence of edge states by the analysis of eigenstates of a finite crystal. Finally, the theoretical predictions of the band structure and the existence of topologically protected edge states are confirmed by numerical simulations of an elastic experimental platform.

\section{Tight-binding of a polymer red phosphorus molecule}
\begin{figure}
    \centering
    \begin{tabular}{c}
    (a)\qquad \qquad \qquad White phosphorus \qquad \qquad \qquad \qquad\\
    \includegraphics[width=0.4\linewidth]{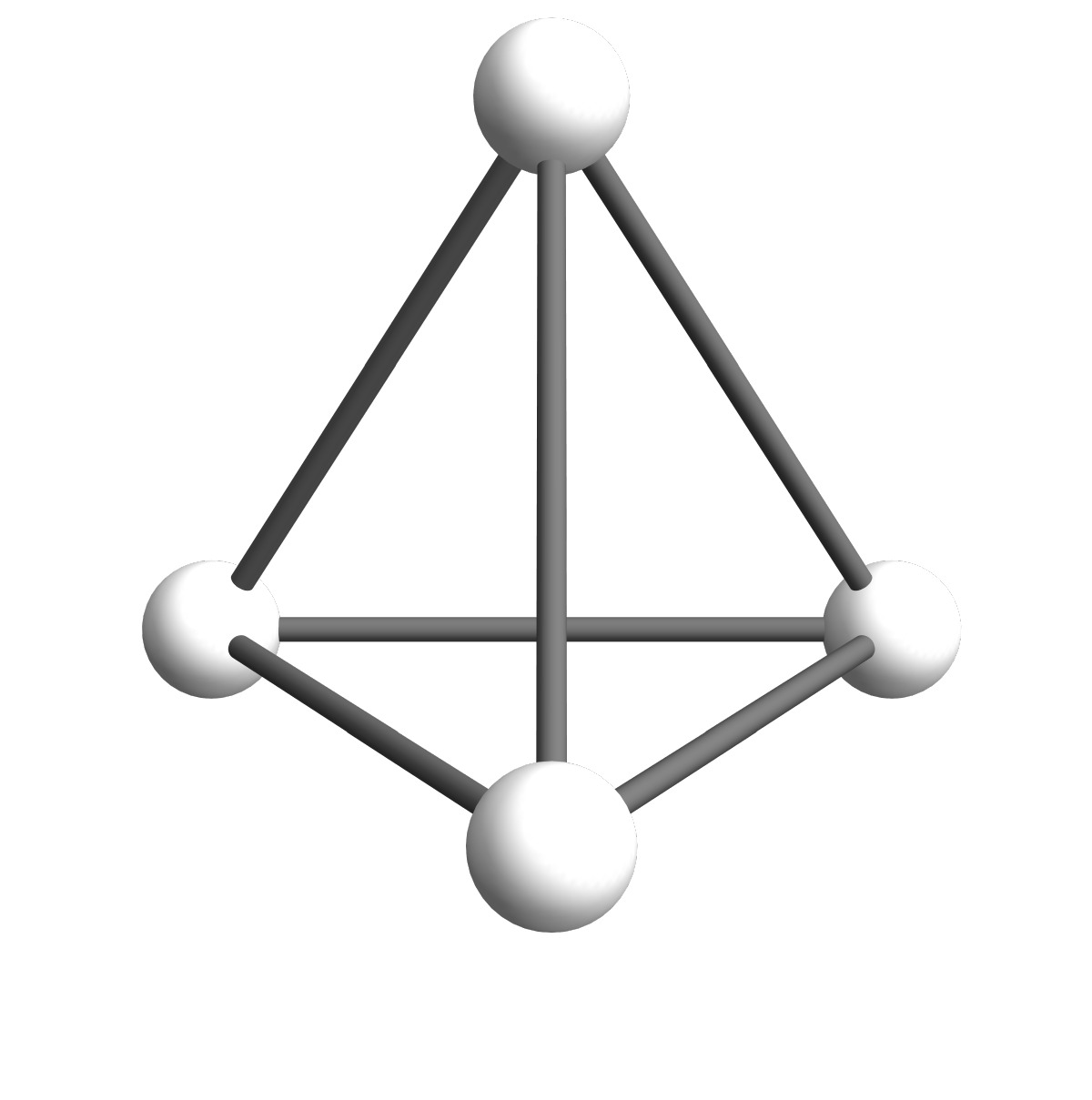}\\
    (b)\qquad \qquad \qquad Red phosphorus \qquad \qquad \qquad \qquad\\
    \includegraphics[width=1\linewidth]{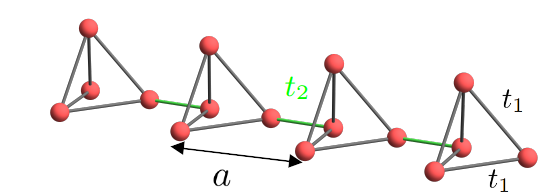}
    \end{tabular}
    \caption{(a) Pictorial representation of the P$_4$ molecule also known as white phosphorus. (b) Red phosphorus lattice formed by a series of open-bond P$_4$ molecules, where $t_1$ and $t_2$ are the intra and inter hopping parameters, respectively.}
    \label{fig:Red_Phosp}
\end{figure}

\begin{figure*}[t!!]
    \begin{tabular}{cccc}
    \centering
    \includegraphics[width=0.24\linewidth]{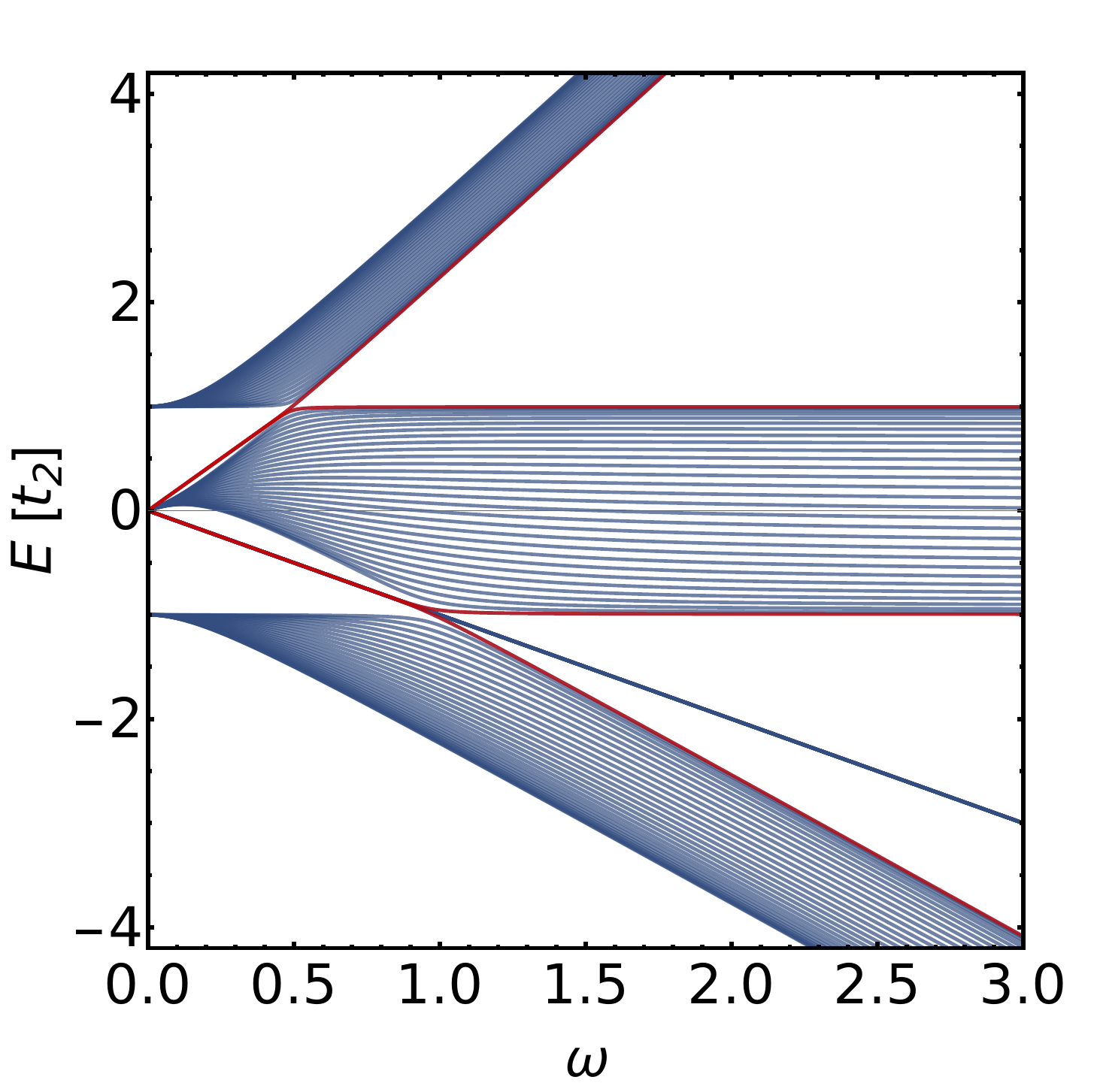}&
    \includegraphics[width=0.25\linewidth]{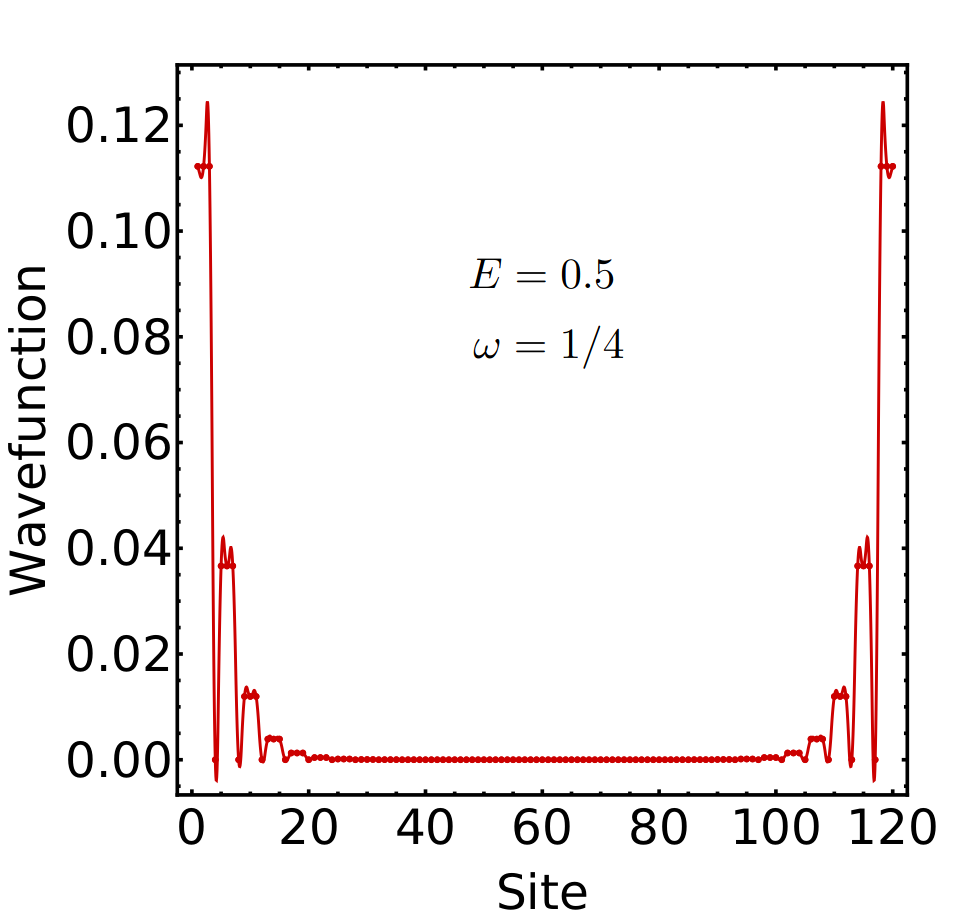}&
    \includegraphics[width=0.25\linewidth]{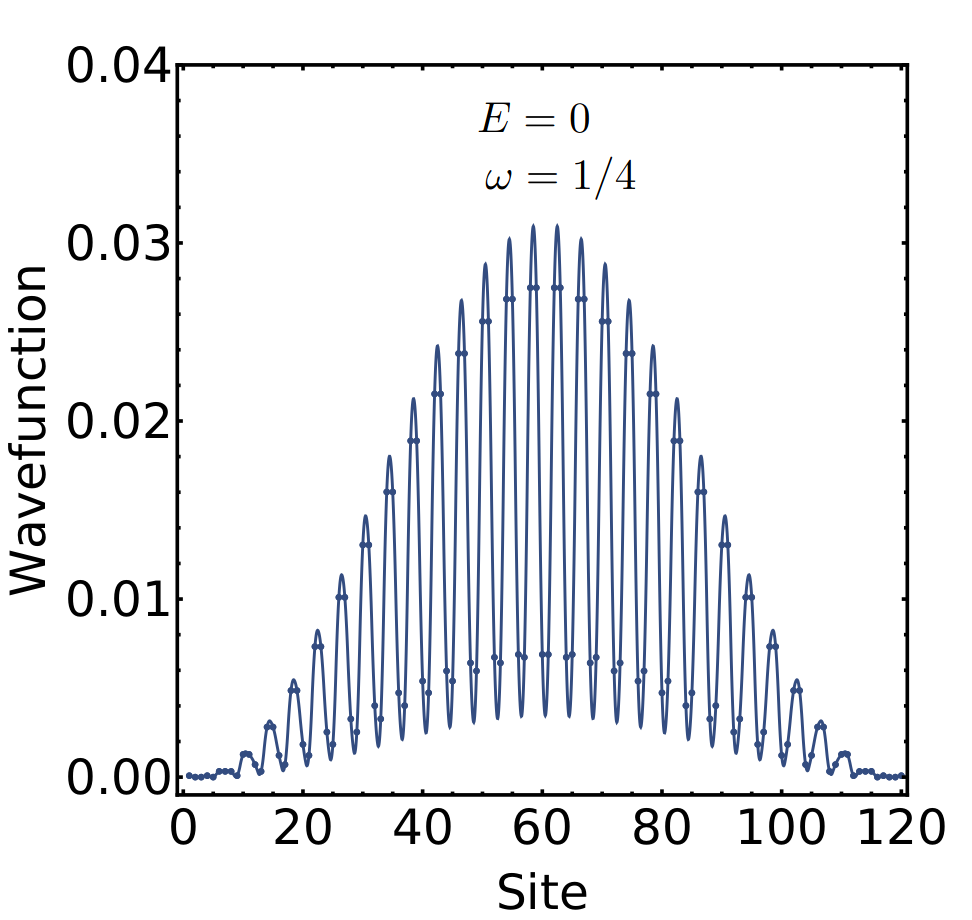}&
    \includegraphics[width=0.25\linewidth]{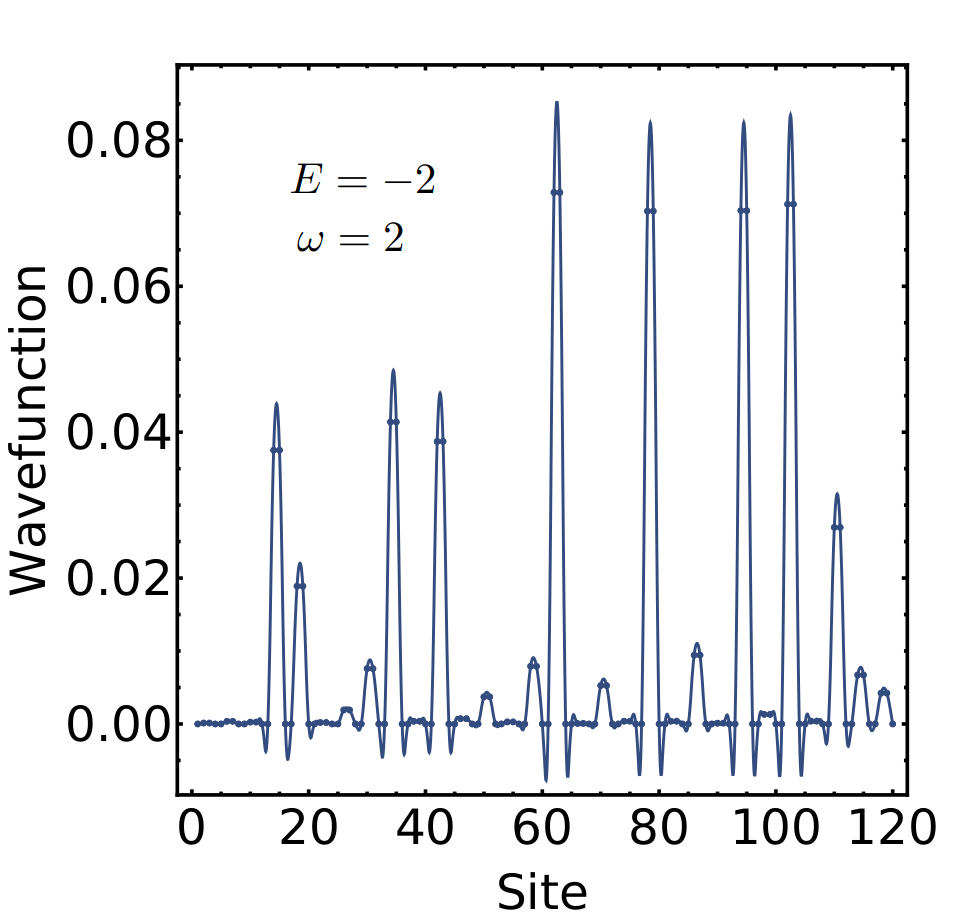}
    \end{tabular}
    \caption{(a) Energy spectrum of the finite red phosphorus chain as a function of $\omega = t_1/t_2$, we fix $t_2 = 1$. The blue bands correspond to bulk states, while the red ones correspond to edge states. (b)-(d) Squared modulus of the wave function for the topological edge (red) and bulk states (blue) as a function of the site number. In (d) is shown the wave function for the unique blue flat band in (a) for non-dispersive bulk states.}
    \label{fig:topredph}
\end{figure*}
Red phosphorus exhibits several allotropes \cite{Zhou2022,Bai2024,Walters2025}, typically found in amorphous form in nature. In this work, however, we consider a hypothetical crystalline one-dimensional structure constructed from P$_4$ molecule units. The P$_4$ molecules, also known as white phosphorus, has a pyramidal geometry with phosphorus atoms located at the vertices, as shown in Fig. \ref{fig:Red_Phosp}(a). Our proposal consists of a periodic chain to explore the topological properties through artificial systems. In Fig. \ref{fig:Red_Phosp}(b) is shown that red phosphorus has a unit cell with an open bond P$_4$ molecule. The chain is formed by connecting neighboring P$_4$ units through an open bond, such that one phosphorus atom of each molecule binds to the adjacent unit. 

Within the tight-binding approach, the three-dimensional geometry of red phosphorus is not essential. Thus, only the projected positions of the atoms along the periodic direction are taken into account. When the Bloch theorem is applied, the dimensionality is indicated through the direction in which the system is periodic.  In Fig. \ref{fig:Red_Phosp}, the hopping parameters $t_1$ and $t_2$ represent the probability amplitude of an electron to hop within and outside the P$_4$ molecule, respectively.

If we consider a finite red phosphorus chain, the tight-binding Hamiltonian is depicted by the following block-diagonal submatrix

\begin{equation}
    h_\textrm{rp}(t_1) = \left(\begin{array}{cccc}
         E_0 & t_1 & t_1 & 0  \\
         t_1 & E_0 & t_1 & t_1  \\
         t_1 & t & E_0 & t _1 \\
         0 & t_1 & t_1 & E_0  
    \end{array}\right),
\end{equation}

\noindent that corresponds to the Hamiltonian of the P$_4$ molecule, and the non-diagonal submatrix, which is given by 

\begin{equation}
    C(t_2) = \left(\begin{array}{cccc}
         0 & 0 & 0 & 0  \\
         0 & 0 & 0 & 0  \\
         0 & 0 & 0 & 0  \\
         t_2 & 0 & 0 & 0  
    \end{array}\right),
\end{equation}

\noindent couples all the neighbor P$_4$ molecules. The full Hamiltonian for the polymer molecule of red phosphorus, in Toeplitz matrix form, is 

\begin{align}
\label{eq:Nssh}
    H_\textrm{fc}(t,t') = \qquad \qquad \qquad \qquad \qquad \qquad \qquad \qquad \qquad &\nonumber\\
    \left(\begin{array}{c c c c c}
        h_\textrm{rp}(t_1)  & C(t_2) & 0 & \cdots & 0  \\
         C^\dagger(t_2) & h_\textrm{rp}(t_1) & C(t_2)  & 0 & \cdot\\
         0 & C^\dagger(t_2) & \cdot & \cdot & \cdot \\
         \vdots & \vdots &  \vdots & \vdots & C(t_2) \\
         0 & 0 & 0 & C^\dagger(t_2) & h_\textrm{rp}(t_1)
    \end{array}\right).&
\end{align}

The diagonalization of this Hamiltonian leads to the energy spectrum, as shown in Fig. \ref{fig:topredph} (a). By considering 30 unit cells for the finite chain, we get 120 eigenenergies as a function of the hopping parameter ratio $\omega \equiv t_1/t_2$, where four energies correspond to edge states (red curves). Rather unconvetionally, the edge states become topological at two different energies in the spectrum. 
Increasing $\omega$, there are topological edge states of energy $E=2t_1$ for the range $t_1<\frac{t_2}{2}$. The second set of topological edge states exists for the range $t_1 < t_2$ and has energy $E=-t_1$, see the corresponding red lines in Fig. \ref{fig:topredph}(a). 
The eigenstates of the Hamiltonian in Eq. \eqref{eq:Nssh} evidence the localization of electrons in the edges for the energies that correspond to topological phases in Fig. \ref{fig:topredph}(b) and bulk states in Fig. \ref{fig:topredph}(c) and (d). It is important to note that there is a constant blue line for the energy $E = -t_1$ in the whole range of $t_1$. This blue line is for non-dispersive bulk states, as seen in Fig. \ref{fig:topredph} (d). 

\section{Equivalence to SSH-type trimer chain: topological band-edge states}

From the finite red phosphorus Hamiltonian in Eq. \eqref{eq:Nssh}, we can derive the Bloch Hamiltonian of the infinite chain applying the Bloch theorem. In this way, the wavefunction in matrix representation is given by 

\begin{eqnarray}\label{BWF}
    \vec{\Psi}(k) & = & (\ldots,A_1\textrm{e}^{ika},A_2\textrm{e}^{ika},A_3\textrm{e}^{ika},A_4\textrm{e}^{ika},\nonumber\\
    & & A_1\textrm{e}^{2ika},A_2\textrm{e}^{2ika},A_3\textrm{e}^{2ika},A_4\textrm{e}^{2ika}, \dots),
\end{eqnarray}

\noindent where $A_j$, with $j = 1,2,3$ and 4, are the amplitudes for the four atomic orbitals in the unit cell and $a$ is the unit cell length. From Schrödinger equation $H_\textrm{fc}(t_1,t_2)\vec{\Psi}(k) = E\vec{\Psi}(k)$, using the Hamiltonian in Eq. \eqref{eq:Nssh}, and wavefunction \eqref{BWF}, the $4 \times 4$ Bloch Hamiltonian yields  
\begin{equation}
    H_\textrm{RP}(k)=\begin{pmatrix}
        0 & t_1 & t_1 & t_2 e^{-iak} \\
        t_1  & 0 & t_1 & t_1  \\ 
        t_1  & t_1 & 0 & t_1  \\
        t_2 e^{iak} & t_1  & t_1  & 0
    \end{pmatrix}.\label{H0}
\end{equation}

\noindent Let us make the following unitary transformation of the Hamiltonian $H_\textrm{RP}$,
\begin{align}\label{H2}
H(k)=UH_\textrm{RP}(k)U^{\dagger}=\begin{pmatrix}
0&\sqrt{2}t_1&t_2e^{-i k a}&0\\
\sqrt{2}t_1&t_1&\sqrt{2}t_1&0\\
t_2e^{i k a}&\sqrt{2}t_1&0&0\\
0&0&0&-t_1
\end{pmatrix},\nonumber\\
\textrm{with} \qquad U=\begin{pmatrix}
1&0&0&0\\
0&\frac{1}{\sqrt{2}}&\frac{1}{\sqrt{2}}&0\\
0&0&0&1\\
0&\frac{1}{\sqrt{2}}&-\frac{1}{\sqrt{2}}&0
\end{pmatrix}.
\end{align}
The Hamiltonian gets a block-diagonal form.
It explicitly reveals decoupling of the flat-band dynamics from the rest of the system.
The states given by multiples of the vector $(0,0,0,1)^t$ correspond to the flat-band energy $-t_1$.
 
\begin{figure}[t!!]
    \centering
    \includegraphics[width=1\linewidth]{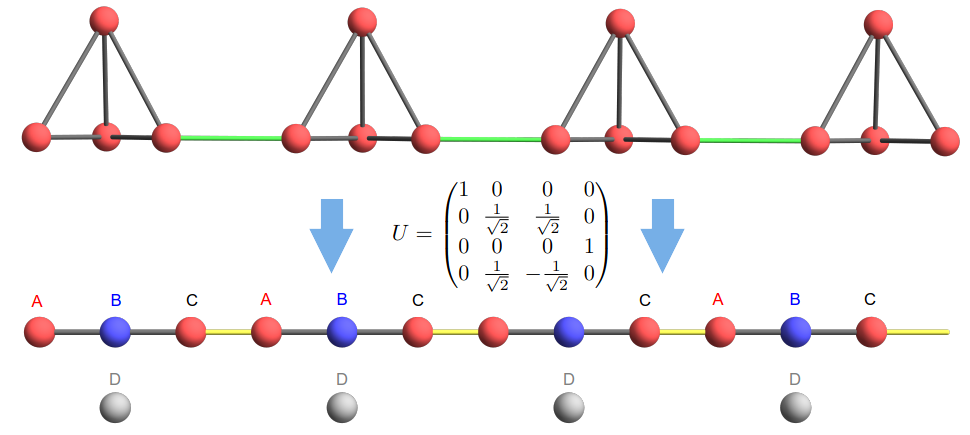}
    \caption{Auxiliary atomic chain described by the Hamiltonian (\ref{H3}). Flat-band states are located on the atoms of the $D$-chain.}
    \label{auxiliarychain}
\end{figure}

The operator (\ref{H2}) can be considered as the Bloch Hamiltonian of an auxiliary atomic ladder composed of two parallel chains of atoms. The first one is an SSH-type trimer chain. Let us call the three atoms in the elementary cell of the chain as $A$, $B$, and $C-$atoms.   The second parallel chain is composed of non-interacting $D$ atoms, see Fig. \ref{auxiliarychain}. The quasi-particles located on these chains are represented by the corresponding components of the transformed Bloch wave function,
$$ \Psi(k)=\begin{pmatrix}{\psi}_A (k),{\psi}_B(k),{\psi}_C(k),{\psi}_D(k)\end{pmatrix}^t.$$
The states located on the $D$-atoms correspond to the flat-band and do not interact with the rest of the system. 
The $B$ atoms have onsite energy $t_1$. 
The tight-binding Hamiltonian for the auxiliary SSH-type trimer associated with the Bloch operator (\ref{H2}) can be written as follows
\begin{eqnarray}\label{H3}
H & = & t_1\sum_{n}B_n^\dagger B_n+\sum_{n}\left(\sqrt{2}t_1 A_n^\dagger B_{n}+\sqrt{2}t_1 B_n^\dagger C_{n}+\right.\nonumber\\
&& \left.t_2A_n^\dagger C_{n+1}+h.c.\right)-t_1 \sum_{n}D_n^\dagger D_n.
\end{eqnarray}

In the following section, we will investigate spectral and topological properties of the system described by $H(k)$ in Eq. \eqref{H2}. As it is unitarily equivalent to $H_\textrm{RP}(k)$, the results will be directly applicable to the chain of red phosphorus. 

\section{Energy bands, Berry phase and inert band-edge states}

The eigenvalues of $H(k)$ in Eq. \eqref{H2} can be found straightforwardly. The secular equation $\det(H(k)-E)=0$ is of fourth order in $E$. As we noted before, one solution corresponds to the flat band energy $E_{fb}=-\omega t_2=-t_1$. The other three energy bands are solutions of a cubic equation for $\epsilon$. The equation  can be written in the following form
\begin{equation}\label{implicitDispersion0}
    \cos{k a}=\frac{1}{4\omega^2}(\epsilon-\omega)(\epsilon^2-1)-\epsilon,\quad \epsilon=E/t_2,\quad \omega=t_1/t_2.
\end{equation} 
The third order equation can be solved analytically with the use of Cardano's formula, see e.g. Appendix in \cite{Jakubsky2022}. This way, we get

\begin{eqnarray}
E_1&= &\frac{t_2}{3} \left[ \omega + 2 \sqrt{3 + 13 \omega^2} \cos\left( \frac{1}{3} \arccos q \right) \right],\nonumber\\
E_2&= &\frac{t_2}{3} \left[ \omega + 2 \sqrt{3 + 13 \omega^2} \cos\left( \frac{2\pi}{3} + \frac{1}{3}\arccos q\right)  \right],\nonumber\\
E_3 & = &\frac{t_2}{3} \left[ \omega - 2 \sqrt{3 + 13 \omega^2} \sin\left( \frac{1}{3} \arcsin q \right) \right],\nonumber\\
q& =&\frac{\omega \left[ -9 + 19 \omega^2 + 54 \omega \cos(k a) \right]}{\left(3 + 13 \omega^2 \right)^{3/2}}.
\end{eqnarray}

\begin{figure*}[t!]
    \centering
    \includegraphics[width=.19\linewidth,trim=1.32cm .95cm 0cm 0cm, clip]{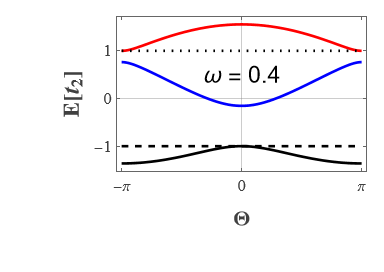}       
    \includegraphics[width=.19\linewidth,trim=1.32cm .95cm 0cm 0cm, clip]{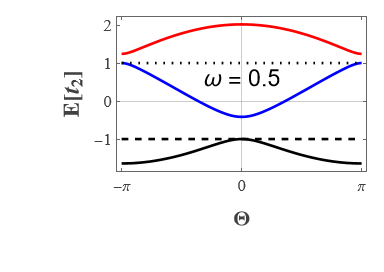}    
    \includegraphics[width=.19\linewidth,trim=1.32cm .95cm 0cm 0cm, clip]{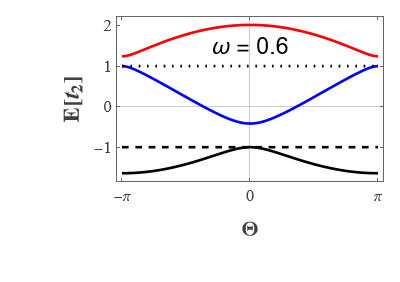}    
    \includegraphics[width=.19\linewidth,trim=1.32cm .95cm 0cm 0cm, clip]{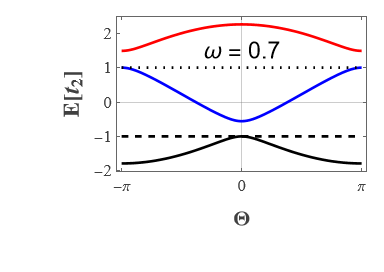}    
    \includegraphics[width=.19\linewidth,trim=1.32cm .95cm 0cm 0cm, clip]{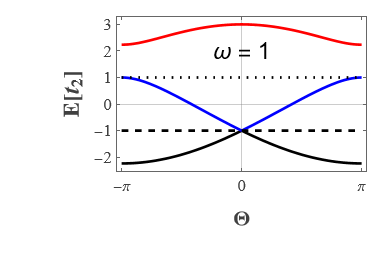}    
 
    \caption{Energy bands $E_{1,2,3}/t_2$ in dependence on $\Theta=k a$ for five different values of $\omega=t_1/t_2$. We fixed $a=1$ and $t_2=1$.}
   \label{bands}
\end{figure*}

The figures of the energy bands are plotted for different values of parameters in Fig. \ref{bands}. 

Let us analyze minima and maxima of the energy bands $E_{1,2,3}$ in the first Brillouin zone. 
The right-hand side of (\ref{implicitDispersion0}) is a third-order polynomial in $\epsilon$ that goes to $\pm \infty$ as $\epsilon\rightarrow \pm\infty$, respectively. The left-hand side can acquire values in the interval $[-1,1]$. Therefore, the solutions of (\ref{implicitDispersion0}) are only the values of $\epsilon$ where the polynomial falls into the interval $[-1,1]$. 
It is straightforward to find their explicit values as they correspond to the solutions of (\ref{implicitDispersion0}) when either $\cos k a =1$ (that is, $k=0$) or $\cos ka =-1$ (that is, $k=\pm\pi/a$). When $k=0$, the equation (\ref{implicitDispersion0}) can be factorized as 
\begin{equation}\label{ebe1}
(\epsilon+1)(\epsilon^2-\epsilon(1+\omega)-\omega (1-4\omega))=0,\end{equation}
whereas for $k=\pm\pi/a$ we have
\begin{equation}\label{ebe2}
(\epsilon-1)(\epsilon^2+\epsilon(1-\omega)-\omega (1+4\omega))=0.\end{equation}

\begin{figure}[t!!]
    \centering
\includegraphics[width=0.9\linewidth]{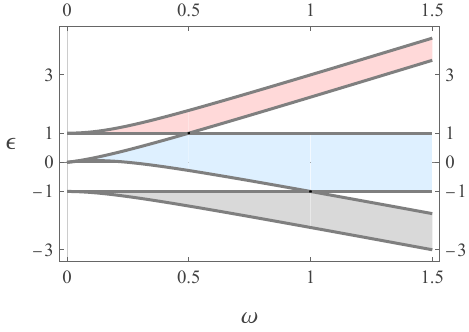}
        \caption{Energy bands in dependence on $\omega=t_1/t_2$. The gap between the lower and middle band gets closed for $\omega=1$, whereas the gap between the second and the third band closes for $\omega=1/2$. The edges of the energy gap are solutions of (\ref{ebe1})  and (\ref{ebe2}).}\label{gaps0}
\end{figure}

It is rather remarkable that one solution in Eqs. \eqref{ebe1}and \eqref{ebe2} is independent on $\omega$,
\begin{equation}
\epsilon_{\pm}=\pm 1\Longrightarrow E_{\pm}=\pm t_2.
\end{equation}
We call $E_{\pm}$ \textit{inert band edges}.  It also follows from (\ref{ebe1}) and (\ref{ebe2}) that two energy bands get closed when the ratio $\omega=t_1/t_2$ is either $\omega=\frac{1}{2}$ or $\omega=1$. The dependence of the energy bands on $\omega$ is illustrated in Fig. \ref{gaps0}.
The eigenstates of $H(0)$ and $H(\pi/a)$ corresponding to the inert band edges $E_{\pm}$ can be found in straightforward manner. The two states coincide, they have exactly the following form,
\begin{align}
\Psi_{inert}&=\begin{pmatrix}1\\0\\-1\end{pmatrix},\nonumber\\
(H(0)+E_-)\Psi_{inert}&=(H(\pi/a)-E_+)\Psi_{inert}=0.
\label{inert}\end{align}
They are located on the sublattices A and C. 
As they do not depend on $\omega$, we call them \textit{inert band-edge states}. 

Let us consider $\omega$ as the tunable parameter whereas  $t_2$ will be fixed. The band gaps close and open again as $\omega$ crosses the  critical values $\omega= 1/2$ and $\omega= 1$. The closing and opening of the energy gaps can be associated with an abrupt change of the topological properties of the energy bands. We calculate the Berry phase as the indicator of non-trivial topological characteristics.

\begin{figure}[t!!]
    \centering
\includegraphics[width=.33\linewidth]{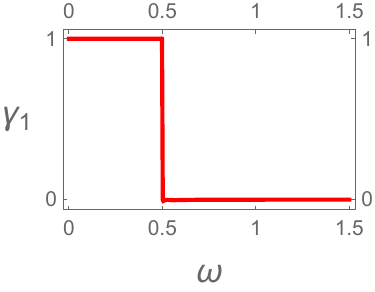}\includegraphics[width=.33\linewidth]{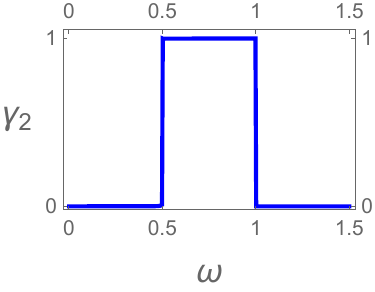}\includegraphics[width=.33\linewidth]{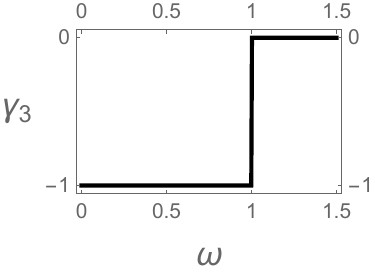}
        \caption{From the left to the right: $\gamma_1$, $\gamma_2$, $\gamma_3$ as functions of $\omega$.  In all cases $t_2=1$.}\label{Berry0}
\end{figure}

The Berry phase can be calculated with the use of the normalized eigenstates of $H(k)$ corresponding to $E_j(k)\equiv t_2\epsilon_j(k)$. They can be found in a rather straightforward manner. They satisfy
\begin{align}
H(k)\Psi_j(k)=E_j(k)\Psi_j(k), \quad \langle \Psi_{j}(k)|\Psi_{j}(k)\rangle=1,\nonumber\\
j=1,2,3.
\end{align} 
We do not present their explicit form here, as the formulas are rather large. The eigenstates can be used to calculate the corresponding Berry phase $\gamma_j$ associated with each $\Psi_j(k)$,
\begin{equation}\gamma_j=-\frac{i}{{\pi}}\int_{-\frac{\pi}{a}}^{\frac{\pi}{a}}\Psi_j^\dagger(k)\partial_{k}\Psi_{j}(k)dk,\quad j=1,2,3.
\end{equation}
The numerical values of $\gamma_j$ as functions of $\omega$ and $t_2$ are illustrated in Fig.\ref{Berry0},

The results confirm that when $\omega$ crosses its critical values, the energy gap closes and the Berry phase of the two touching bands undergoes a corresponding change. In this process, the associated topological charge is transferred from one band to the other. The presence or absence of inert band edges correlates directly with the Berry phase: bands with zero or two inert edges are topologically trivial, whereas a band hosting a single inert edge acquires a non‑trivial Berry phase. This intriguing relation between topological phase and inert edge states merits a more detailed analysis elsewhere. The resulting topological characteristics of the bands as functions of $\omega$ are summarized in Tab.~\ref{tab:gamma-values}.

\begin{table}[h!]
\centering
\renewcommand{\arraystretch}{1.3}
\begin{tabular}{|c|c|c|c|}
\hline
 & $\omega < \frac{1}{2}$ & $\frac{1}{2} < \omega < 1$ & $1 < \omega$ \\
\hline
$\gamma_1$ & $1$ & $0$ & $0$ \\
\hline
$\gamma_2$ & $0$ & $1$ & $0$ \\
\hline
$\gamma_3$ & $-1 $ & $-1$ & $0$ \\
\hline
\end{tabular}
\caption{The values of the Berry phase $\gamma_{1,2,3}$ for each energy band in dependence on $\omega$.}
\label{tab:gamma-values}
\end{table}

\section{Artificial red phosphorus based on coupled-resonator phononic metamaterial}

\begin{figure}[t!]
\centering
\includegraphics[width=0.8\linewidth]{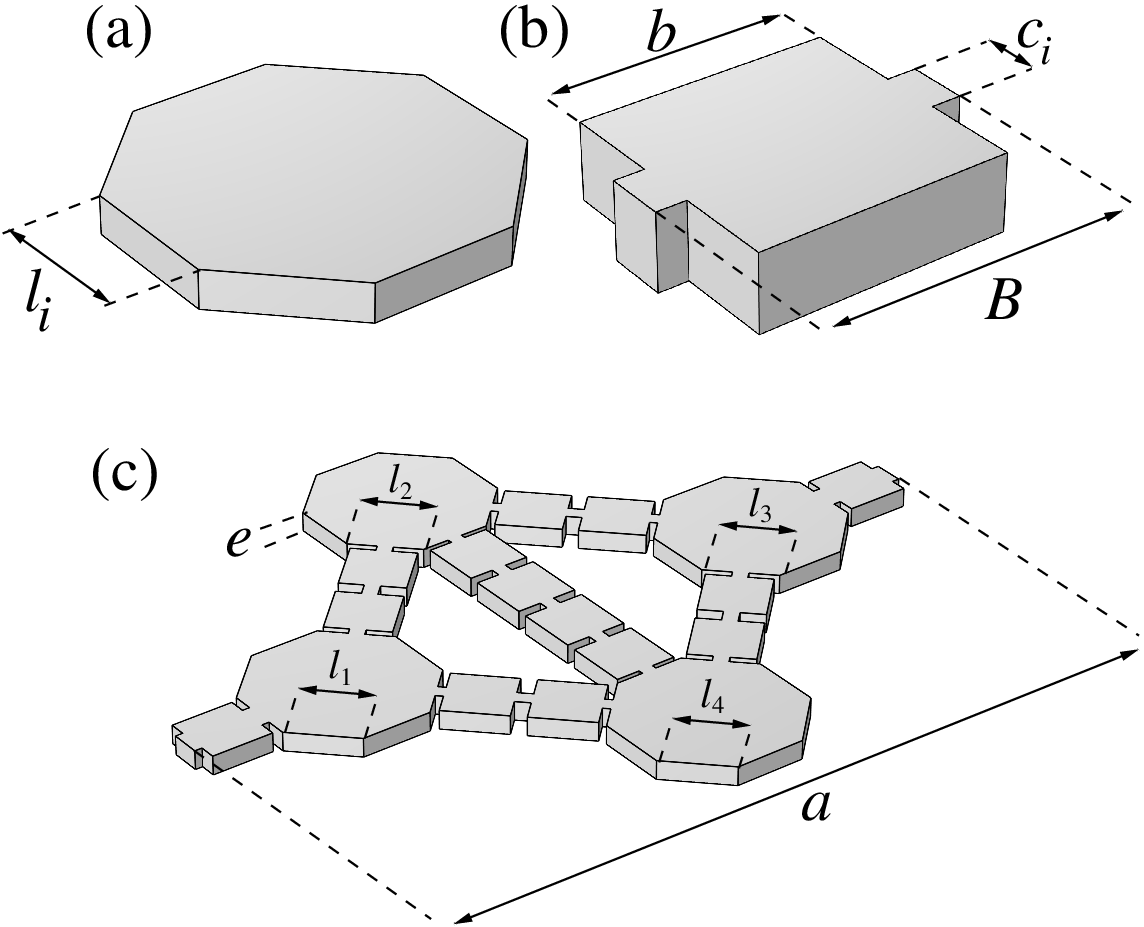}
\caption{(a)  Atomic artificial site composed of an octagonal-shaped resonator of length $l$. The bond is a finite phononic crystal whose unit cell id given in (b); it is composed of a square plate of length $b$ and two smaller plates with length $B$-$b$ and width $c_i$. 
Here  $c_1$ controls the hopping between intracell sites and $c_2$ the intercell hopping. (c) Unit cell (of length  $a$) of the artificial red phosphorous designed on a plate of thickness $e$. The geometrical parameters are $l_1 =l_3= 23$~mm and $l_2=l_4=23.1$~mm, $B$ = 26~mm, $b$ = 21~mm, $c_1$ = 7, 6.652, 6.4, 6 and 5.5~mm, $c_2$= 8.6258~mm, $e$ = 6.35~mm and $a$ = 26.01~mm. The mechanical properties of the plates are those of aluminum.}
\label{PhononicUnitCell}
\end{figure}

One possible platform to emulate the hypothetical red phosphorus molecule shown in Fig. \ref{fig:Red_Phosp} (b) is provided by phononic crystals. 
These artificial structures enable precise control of the hopping parameters by tuning the geometry of the waveguides that evanescently couple the resonances of the meta-atoms. 
Fig.~\ref{PhononicUnitCell} shows the unit cell designed to emulate red phosphorus. 
It is composed of octagonal resonators, Fig.~\ref{PhononicUnitCell} (a), coupled via finite phononic crystals (FPnCs). 
The FPnCs unit cell, see Fig.~\ref{PhononicUnitCell} (b), consists of a square plate with two attached cuboids. 
As shown in Fig. ~\ref{PhononicUnitCell}, two types of FPnCs are employed: FPnC-A, with two and four unit cells to implement intracell hopping amplitudes, and FPnC-B, with two unit cells to realize intercell hopping. 
As discussed below, when a resonance of the octagonal plates lies within the bandgap of both FPnCs, the corresponding vibrational modes become spatially confined to the resonators. 
These localized modes are weakly coupled through evanescent Bloch waves that decay inside the FPnCs, giving rise to an effective tight-binding phononic regime. 
Two almost equal sizes of the octagonal plates were used to compensate for small shifts (of a few hertz) in the resonance frequencies caused by the different coupler configurations.

\begin{figure}[t!]
    \centering
    \includegraphics[width=\linewidth]{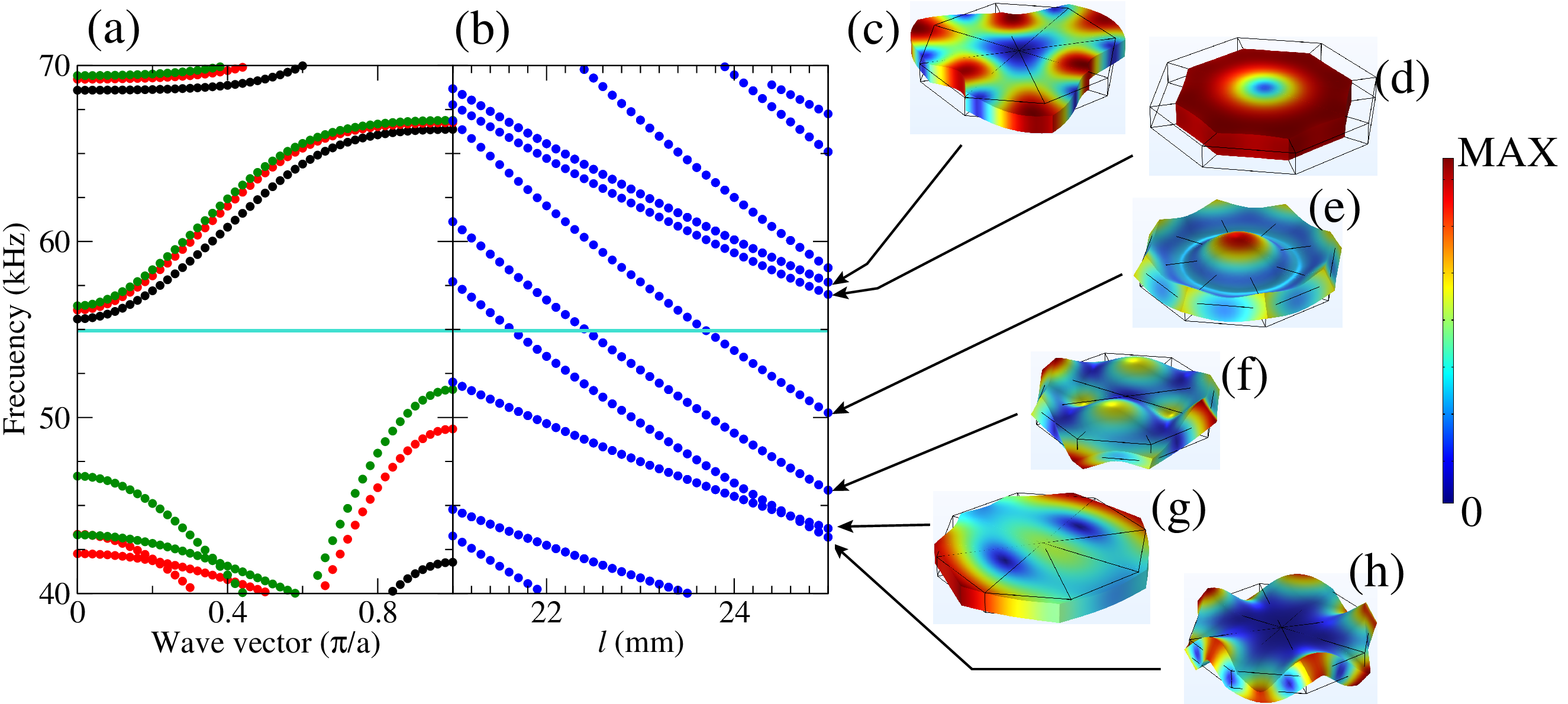}
    \caption{(a) Band structure of a crystal whose unit cell is given in Fig.~\ref{PhononicUnitCell} (b). The dots green, red, and black correspond to different values of $c_2 = 3.5$~mm, 5.4~mm, and 6~mm. (b) Frequency spectrum of an octagonal resonator as a function of its length $l$. The horizontal (cyan) line corresponds to $55$~kHz. (c) Normal-mode wave amplitudes of the resonator for $l=25$~mm above $40$~kHz with increasing frequency. (d-h) Other normal-mode wave amplitudes of the resonator near the bandgap in (a).}
\label{ResonatorAndFPnCs}
\end{figure}

\begin{figure*}[t!]
\centering
\includegraphics[width=\linewidth]{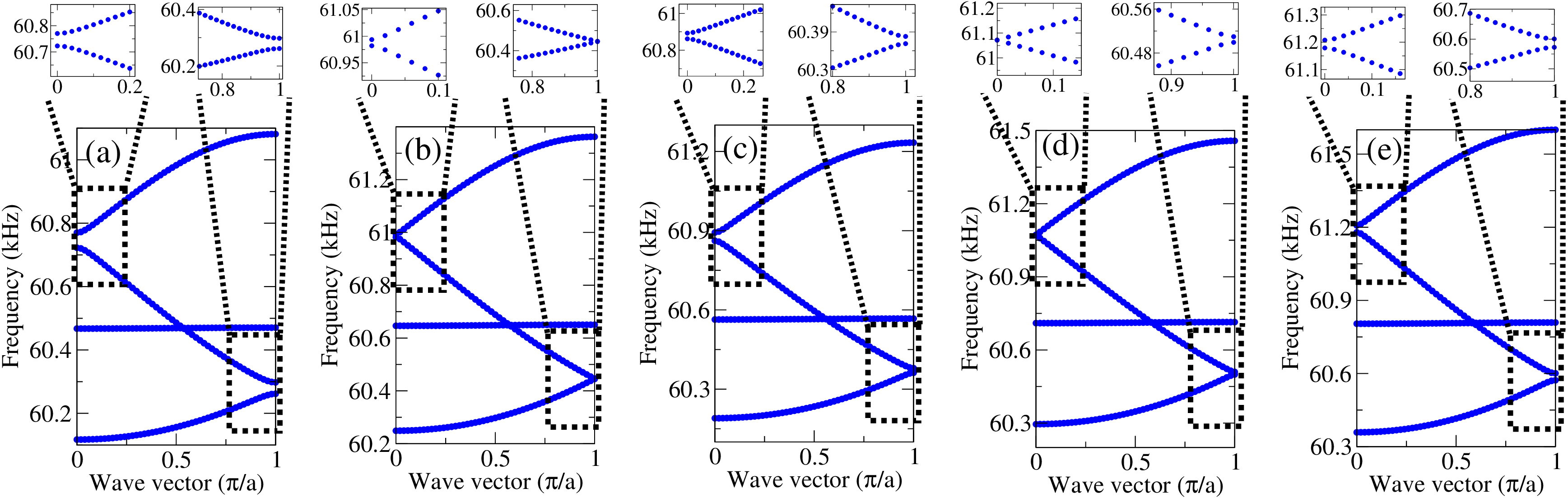}
\caption{Band structure of the artificial red phosphorus 1D chain. $c_2$ = 8.625 mm and $c_1$ = 7, 6.652, 6.4, 6, and 5.5~mm for cases (a) to (e), respectively. This corresponds to $t_1$ varying from 
$t_2\approx 0.25 t_1$ to $t_2\approx 1.5 t_1$, scanning the two topological transitions.}
\label{PhononicBS}
\end{figure*}

\begin{figure*}[t!]
\centering
\includegraphics[width=0.9\textwidth]{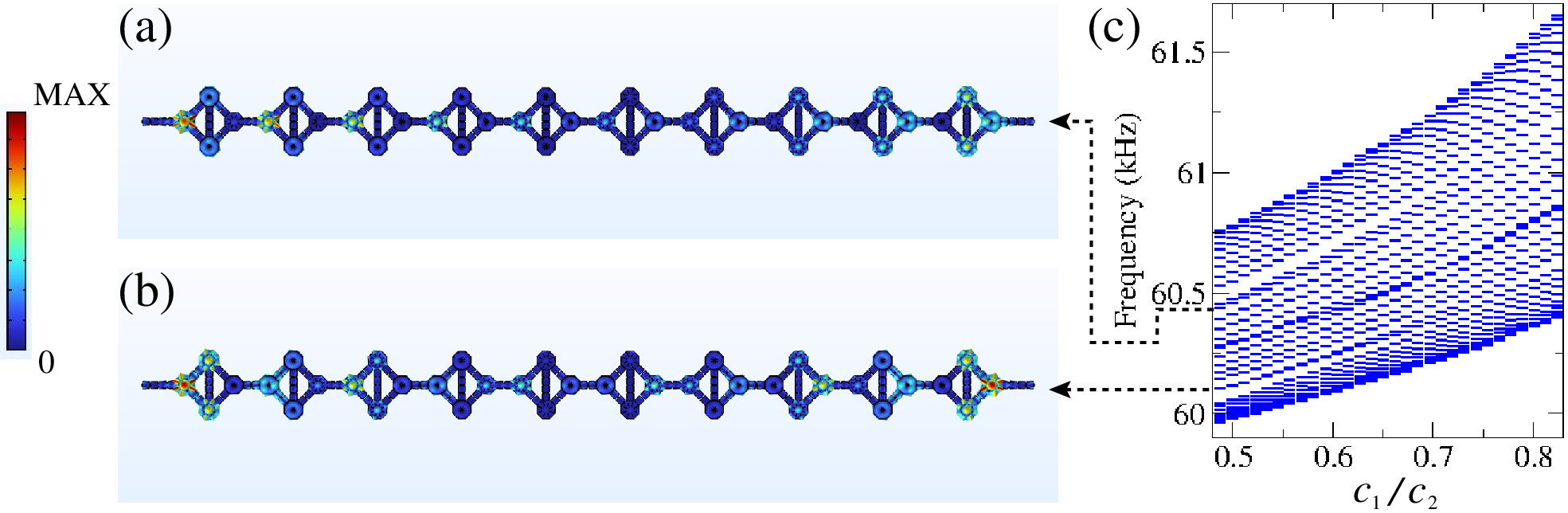}
\caption{Topologically protected states (soliton edge states) of a finite artificial phononic red phosphorous at different frequencies. In (a), the antisymmetric wave amplitude is shown, while in (b) the symmetric one is depicted, both decay evanescently toward the center of the chain. (c) Frequency spectrum as a function of $c_1$ associated to the intracell hopping amplitude $t_1$. The intercell hopping amplitude $t_2$ is fixed by $c_2 = 8.625$~mm. The occurrence of a double topological phase transition at different frequencies is identified.}
\label{transition}
\end{figure*}

All numerical results presented below were calculated using finite elements (COMSOL Multiphysics). 
In Fig.~\ref{ResonatorAndFPnCs}~(a), the band structures of the finite phononic crystals, used to emulate the different hoppings, are given; they were calculated using Floquet boundary conditions. 
Since the intracell hopping $t_1$ is emulated with FPnCs of two and four cells, see Fig.~\ref{PhononicUnitCell}, two different values of $c_1$ were used to give the same hopping parameter; $c_1=5.4$~mm for a coupler with two cells and $c_1'=5.41$~mm for the coupler with four cells. 
As can be seen, for the parameters used, the band structures present a bandgap around 55~kHz. 
The frequency of the resonator will be located within this bandgap. 
In Fig.~\ref{ResonatorAndFPnCs}~(b), the normal-mode frequencies of the octagonal plate are given as a function of its side length $l$. 
A frequency of the octagonal plate falls within the bandgap of the FPn\-Cs for $l\approx23.75$~mm. 
The out-of-plane normal mode wave amplitude associated with this frequency is given in Fig.~\ref{ResonatorAndFPnCs} (e) among with other wave amplitudes. 

The band structures obtained for the artificial red phosphorus are given in Fig.~\ref{PhononicBS} for different values of the parameter $c_1$ associated with the intracell hopping $t_1$; the parameter $c_2$ associated to the intercell hopping $t_2$ is maintained constant. 
The geometrical parameters of the unit cell are given in the caption of Fig.~\ref{PhononicUnitCell}. 
Since the unit cell has four resonators, four bands appear. 
One flat band appears and corresponds to modes moving in the Y direction, {\it i.e.}, without transport in the X direction; this flat band is then uncoupled from the other bands and can be classified as an orthogonal flat band\cite{Danieli2024}. 
As it can be seen, also  in Fig.~~\ref{PhononicBS}, the three remaining bands present two topological phase transitions. 
When $c_1\gtrsim7$~mm, all bands are separated by small gaps.
The middle and lower bands join, forming a Dirac cone for $c_1\approx 6.652$~mm; the bands then separate by a small bandgap for $c_1\approx 6.4$~mm. 
For $c_1\approx 6.0$~mm, the middle and upper bands join, forming a Dirac cone. 
For values of $c_1 \lesssim 5.5$~mm, all bands are again separated by small gaps. 
This is in agreement with the theoretical results of the previous section.

A finite coupled-resonator phononic phosphorus of $40$ sites, {\sl i.e.} 10 unit cells,  is now analyzed; its shape can be seen in Fig.~\ref{transition} (a) and (b). 
The frequency spectrum is displayed in Fig.~\ref{transition} (c). 
Here, with a value of $c_1/c_2\approx 0.5$ the topologically protected states appear at $\approx 60.1$~kHz and $\approx60.45$~kHz. 
The topological nature of this states is apparent in this figure since they remain within the bandgaps when varying $c_1/c_2$ for a wide range of parameters.

The wave amplitudes associated to the topologically protected states are given in Fig.~\ref{transition}~(a) and (b) and, as expected, are localized at the borders. 

In (a), an antisymmetric wave amplitude is shown, while in (b) a symmetric one is depicted.
The flat band is also captured by the phononic model and is appears around $60.75\,\mathrm{kHz}$ for $c_1/c_2 = 0.85$. 
This confirms that these are states do not interact with the other bands.
The finite phononic design could be used to experimentally measure the dual energy-differentiated topological transition described here.

\section{Conclusions and final remarks}
We proposed an artificial version of the hypothe\-tical red phosphorus molecule that exhibits two topological transitions emerging in two different energies. This behavior is atypical, as most one-dimensional topological chains, such as the Su-Schrieffer-Heeger model, host edge states pinned at zero energy in the middle of the band gap. Through a tight-binding study of both finite and infinite chains, we identified distinct topological phases emerging for hopping parameter regimes $t_1 \leq \frac{t_2}{2}$ and $t_1 \leq t_2$. 

A key result of this work is the identification of a unitary transformation that block-diagonalizes the Bloch Hamiltonian, separating the dispersive bands from a trivial flat band. This reduction reveals that the red phosphorus lattice can be effectively interpreted as an SSH-type trimer chain. Within this framework, we derived exact analytical expressions for the energy spectrum and eigenstates, enabling a direct and transparent computation of the Berry phase.

The topological characterization of the system unco\-vers an unusual dependence of the Berry phase on the hopping parameter ratio $\omega = t_1/t_2$, with nonzero values in the topological regimes. This behavior departs from that of conventional one-dimensional topological models and highlights the energy-dependent nature of the topological transitions. We observed correlation between existence of the inert band edge states (\ref{inert}) and topological charge of the energy bands that would be worth of further ana\-lysis.

Finally, we demonstrate that phononic crystals are realistic platforms for emulating red phosphorus chains. 
Finite element calculations of vibrational modes and phononic band structures fully corroborate the topological predictions of the tight-binding model. 
These results are experimentally accessible in phononic crystals and metamaterials as those of Refs.~\citep{MartinezArgueello2022,Martinez-EsquivelEtAl,PennecEtAl}, and open new avenues for realizing robust topological states, charge pumping protocols, engineering quantum states, and testing fundamental topological concepts in controllable toy-model platforms.

\section{Acknowledgments}
Y.B.-O. gratefully acknowledges funding from UNAM-PAPIIT IA106223 and IA102125. VJ acknowledges the assistance provided by the Advanced Multiscale Materials for Key Enabling Technologies project, supported by the Ministry of Education, Youth, and Sports of the Czech Republic. Project No. CZ.02.01.01/00/22 008/0004558, Co-funded by the European Union.”AMULET project. B.~Manjarrez-Montañez acknowledges SECIHTI for the financial support of his doctoral studies. R.A.M.-S. was supported by UNAM-PAPIIT under project IN108825. 

\bibliography{RedPhosph}
\end{document}